\documentclass[runningheads]{llncs}
\usepackage{graphicx}
\usepackage{times}
\usepackage{epsfig}
\usepackage{amsmath}
\usepackage{xcolor}
\usepackage{amssymb}
\usepackage{bm}
\usepackage{subfig}
\begin{document}
\title{Synthetic Augmentation and Feature-based Filtering for Improved Cervical Histopathology Image Classification}
\titlerunning{Synthetic Augmentation with Feature-based Filtering}
% If the paper title is too long for the running head, you can set
% an abbreviated paper title here
%
\author{Yuan Xue\inst{1}, Qianying Zhou\inst{1}, Jiarong Ye\inst{2}, L. Rodney Long\inst{3}, Sameer Antani\inst{3}, Carl Cornwell\inst{3}, Zhiyun Xue\inst{3}, and Xiaolei Huang\inst{1}}
\index{Xue, Yuan}\index{Zhou, Qianying}\index{Ye, Jiarong}\index{Long, Rodney}\index{Antani, Sameer}\index{Cornwell, Carl}\index{Xue, Zhiyun}\index{Huang, Xiaolei}

\authorrunning{Y. Xue~\emph{et al.}}
% First names are abbreviated in the running head.
% If there are more than two authors, 'et al.' is used.
%
\institute{ College of Information Sciences and Technology, Pennsylvania State University, University Park, PA, USA
\and
Department of Statistics, Pennsylvania State University, University Park, PA, USA
\and
National Library of Medicine, National Institutes of Health, Bethesda, MD, USA
}
% \institute{Princeton University, Princeton NJ 08544, USA \and
% Springer Heidelberg, Tiergartenstr. 17, 69121 Heidelberg, Germany
% \email{lncs@springer.com}\\
% \url{http://www.springer.com/gp/computer-science/lncs} \and
% ABC Institute, Rupert-Karls-University Heidelberg, Heidelberg, Germany\\
% \email{\{abc,lncs\}@uni-heidelberg.de}}
%
\maketitle              % typeset the header of the contribution
\begin{abstract}
Cervical intraepithelial neoplasia (CIN) grade of histopathology images is a crucial indicator in  cervical biopsy results. Accurate CIN grading of epithelium regions helps pathologists with precancerous lesion diagnosis and treatment planning. Although an automated CIN grading system has been desired, supervised training of such a system would require a large amount of expert annotations, which are expensive and time-consuming to collect. In this paper, we investigate the CIN grade classification problem on segmented epithelium patches. We propose to use conditional Generative Adversarial Networks (cGANs) to expand the limited training dataset,
by synthesizing realistic cervical histopathology images. While the synthetic images are visually appealing, they are not guaranteed to contain meaningful features for data augmentation. To tackle this issue, we propose a synthetic-image filtering mechanism based on the divergence in feature space between generated images and class centroids in order to control the feature quality of selected synthetic images for data augmentation. Our models are evaluated on a cervical histopathology image dataset with limited number of patch-level CIN grade annotations. Extensive experimental results show a significant improvement of classification accuracy from $66.3\%$ to $71.7\%$ using the same ResNet18 baseline classifier after leveraging our cGAN generated images with feature based filtering, which demonstrates the effectiveness of our models.

% We also provide a new cervical histopathology image dataset on the patch level. This new dataset has 2210 patches in total, and 4 classes which are normal, CIN 1, CIN 2, and CIN 3. Each patch has unified size of 256 x 128 pixels.

% \keywords{Cervical histopathology classification  \and Unsupervised data augmentation \and Feature based filtering}
\end{abstract}
%
%
%
% \vspace{-5pt}
\section{Introduction}
% \vspace{-5pt}
% \subsection{A Subsection Sample}
Cervical cancer is the fourth-most frequently diagnosed %and malignant 
cancer among women all over the world~\cite{Bray2018GlobalCS}. 
%Over 80\% cases of death due to cervical cancer occur in the less developed regions of the world, which is mainly because of the lack of resource for early detection~\cite{xu2017multi}. 
The diagnosis of cervical cancer and its precancerous stages can be accomplished through assessment of histopathology slides of cervical tissue by pathologists. 
%through Pap test, by examining the occurrence of abnormal cells through the histopathology slides by pathologists. 
An important outcome of the assessment is the 
%Among all visual examinations and biopsy results of histopathology slides for cervix tissue, 
cervical intraepithelial neoplasia (CIN) grade, an important indicator for abnormality
assessment identified by the abnormal growth of cells on the surface of the cervix. CIN grade can be divided into CIN1, CIN2, and CIN3 with increased severity from mild to severe. Thus, accurate assessment of CIN grade is crucial for diagnosis and treatment planning of cervical cancer.
%For CIN1, dysplastic squamous cells present in one third of the epithelium from the bottom. For CIN2, dysplastic squamous cells may be present in the basal two-thirds of the epithelium. For CIN3, dysplastic squamous cells fill the whole epithelium. 
%In addition to the position of dysplastic squamous cells, pathologists also assess cellular-level features visually, which is difficult due to the large number of nuclei present and the complex visual field~\cite{guo2016nuclei}. 

Considering the shortage of pathologists, an automatic cervical histopathology image classification system has great potential 
in under developed regions for its low cost and accessibility. Moreover, such a system can help pathologists with diagnosis and potentially mitigate the inter- and intra- pathologist variation. 
Existing literature~\cite{chankong2014automatic,guo2016nuclei} have studied various supervised learning methods for nuclei-based cervical cancer classification.
%learning based methods have been applied to automatic classification of cervical histopathology images.~\cite{chankong2014automatic,guo2016nuclei}. 
Chankong~\emph{et al.}~\cite{chankong2014automatic} proposed automatic cervical cancer cell segmentation and classification using fuzzy C-means (FCM) clustering and various types of classifiers.
%including Bayesian classifier, neural networks (NN), linear discriminant analysis (LDA) and support vector machines (SVM). 
Guo~\emph{et al.}~\cite{guo2016nuclei} designed hand-crafted nuclei-based features for fusion-based classification on digitized epithelium histopathology slides with linear discriminant analysis (LDA) and support vector machines (SVM) classifier.  While accomplishments have been achieved with fully-supervised learning methods, they require large amounts of expert annotations of cervical whole slide images. Since the annotation process can be tedious and time-consuming, it often results in limited number of labeled data available for supervised learning models.

Recently, several works have leveraged unsupervised learning methods, more specifically, Generative Adversarial Networks (GANs)~\cite{goodfellow2014generative} in medical image analysis to mitigate the small dataset sizes and limited annotations~\cite{madani2018chest,ren2018adversarial,frid2018gan}. Frid-Adar~\emph{et al.}~\cite{frid2018gan} investigated conditional GANs (cGANs)~\cite{mirza2014conditional} to generate synthetic CT images and improved the performance of CNN in liver lesion classification, by adding generated images into the training data as data augmentation. Similarly, Madani~\emph{et al.}~\cite{madani2018chest} found GAN based data augmentation achieved higher accuracy than traditional augmentation in Chest X-ray classification. Ren~\emph{et al.}~\cite{ren2018adversarial} explored a method to classify the prostate histopathology images by domain adaptation so that knowledge learned in the source domain can be transferred to the target dataset without annotation. 
%A discriminator is trained using GAN loss to learn the domain invariant feature between the source and target dataset, then the knowledge learned in the source dataset can be transferred to the target dataset without annotation. 
% Hu~\emph{et al.}~\cite{hu2018unsupervised} proposed an pipeline for unsupervised cell-level visual representation learning which incorporated category information into the learned features. %BenTaieb~\emph{et al.}~\cite{bentaieb2018adversarial} utilized GANs to solve the staining inconsistencies problem in the histopathology image classification task. 
Although the GANs in previous applications can generate visually appealing synthetic images, the feature quality of generated images varies among examples and not all of them are guaranteed to contain meaningful features to improve the model performance in the original task. 
%To tackle this issue, a feature based filtering mechanism is required to further control the quality of GAN synthetic images.

In this paper, we study the 4-class (normal, CIN 1-3) cervical histopathology image classification problem based on a ResNet18~\cite{he2016deep} baseline classifier. We run and evaluate our models on a heterogeneous epithelium image dataset with limited number of patch-level annotations. Images in the dataset have various color, shapes and texture which makes the classification very challenging. While the capability of the baseline model is limited by the number of training data, we propose a cGAN based image synthesis model to generate high-fidelity synthetic epithelium histopathology patches to expand the training data. To improve the diversity of synthetic images, we incorporate the minibatch discrimination~\cite{salimans2016improved} to reduce the closeness between
examples inside a minibatch. Moreover, unlike previous works which directly added the generated data into the training set, we apply a feature based filtering mechanism to further improve the feature quality of the synthetic images added. We first pre-train a feature extractor using baseline ResNet18 model and calculate feature centroids for each class as the average feature of all training images. The generated images are then filtered based on the distance to the corresponding centriod in the feature space. Experimental results show that our proposed cGAN model along with the feature based filtering significantly outperform the baseline ResNet18 model and the traditional augmentation methods. %During experiments, we carefully examine different components of our model and explore the principles of GAN based data augmentation in medical image analysis.
% Some studies have proved that current neural network has limitation because it might memorize the training data even in the presence of strong regularization. Honeyi~\emph{et al.} proposed a data augmentation method, \textit{mixup}, to linear combine the pairs of examples in the training set and their labels~\cite{mixup}. This method is different from the traditional data augmentation method because it breaks the boundary between categories. In our project, we can also try their method when we are incorporating the generated images into our training set. 
\vspace{-5pt}
\section{Methodology}
\begin{figure}[t]
\begin{center}
% \fbox{\rule{0pt}{2in} \rule{0.9\linewidth}{0pt}}
  \includegraphics[width=0.99\linewidth]{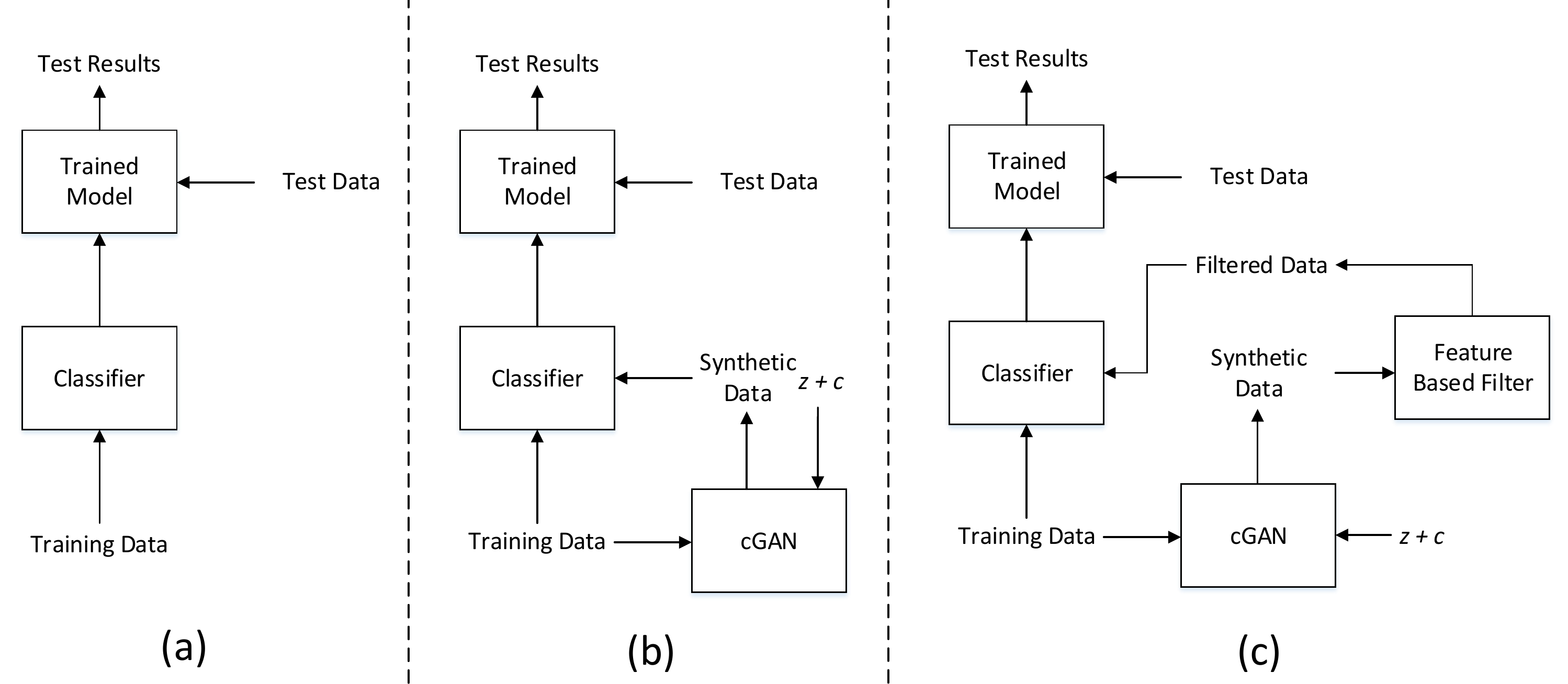}
\end{center}
  \caption{Illustration and comparison between different training processes. (a) Traditional training pipeline; (b) Conditional GAN augmented training pipeline; (c) Our proposed cGAN augmentation with feature based filtering. The input to the cGAN are noise vector $z$ and label condition vector $c$.}
\label{fig:architecture}
% \vspace{-10pt}
\end{figure}

An overall illustration of our proposed data augmentation pipeline can be found in Fig.~\ref{fig:architecture}. In traditional fully-supervised training, the model is trained on training images and the inference is done by feeding the test data to the trained model. In previous GAN-based augmentation works~\cite{frid2018gan,madani2018chest}, a GAN model is first trained to generate some synthetic images based on the training data, then the generated images are added to the original training data as a data augmentation strategy. However, since the discriminator in GAN only outputs a high level judgement ($0$ or $1$) of the fidelity of generated images, such pipeline cannot guarantee that the generated data have similar features to the real images which contribute to the classification task. To tackle this issue, we propose a feature based filtering mechanism to further improve the feature quality and fidelity of the synthetic images. We first introduce the cGAN model used in our framework.
% \vspace{-5pt}
\subsection{Theoretical Preliminaries}
The conventional cGANs~\cite{mirza2014conditional} have an objective function defined as:
\begin{small}
% \begin{equation}
% \min_{\theta_G} \max_{\theta_D} \mathcal{L}_{\text{cGAN}} = \mathbb{E}_{x\sim p_\mathrm{data}}[\log\sigma (D(x,c))] + \mathbb{E}_{z\sim \mathcal{N}}[\log (1 - \sigma(D(G(z,c))))] \enspace .\label{Eq:cGAN}
% \end{equation}

% \textcolor{red}{
\begin{equation}
\min_{\theta_G} \max_{\theta_D} \mathcal{L}_{\text{cGAN}} = \mathbb{E}_{x\sim p_\mathrm{data}}[\log D(x \mid c)] + \mathbb{E}_{z\sim \mathcal{N}}[\log (1 - D(G(z \mid c)))] \enspace .\label{Eq:cGAN}
\end{equation}
% }

\end{small}
In the objective function above,  $\theta_G$ and $\theta_D$ represent the parameters for the generator $G$ and discriminator $D$ in cGAN, respectively. $x$ represents the real data from an unknown distribution $p_\mathrm{data}$ and $c$ is the conditional label (\emph{e.g.}, CIN grades). $z$ is a random vector for the generator $G$, drawn from a normal distribution $\mathcal{N}(0,1)$. $G$ is trained to fool the discriminator with synthetic data by minimizing the objective. Meanwhile, $D$ that takes both $z$ and $c$ as input is trained to maximize the objective, aiming to distinguish real data and synthetic images generated by $G$.

During the experiments, we observe that the intra-class diversity of the generated images greatly affects the data augmentation performance. Consider a mode-collapse GAN model which only generates very limited number of modes, all generated images will look similar. In such case, no matter how many examples are added to the training data, only few of them contribute to the data augmentation. To this end, we incorporate the minibatch discrimination module~\cite{salimans2016improved} into our discriminator to reduce the homogeneity between generated 
examples inside a minibatch. Let $f(x_i) \in R^A$ denote a feature vector for input $x_i$
from an intermediate layer in the discriminator,
where $A$ $=$ out channels $\times$ weight $\times$ height. We multiply
$f(x_i)$ by a transformation matrix $T \in R^{A\times B\times C}$ to compute a matrix $M_i \in R^{B\times C}$, where $B$ represents the number of output features of the minibatch discriminator, and $C$ refers to the number of kernel dimension which is set to 3 by default in our experiments. The similarity of the image $x_i$ with the rest of images $x_j$ in the same batch is computed as
\begin{small}
\begin{equation}
o(x_i)_b = \sum_{j=1}^{n} \exp (-||M_{i,b} - M_{j,b}||_{L_1}) \enspace ,\label{Eq:minibatch}
\end{equation}
\end{small}
and $o(x_i) = [o(x_i)_1, o(x_i)_2, ..., o(x_i)_B] \in R^B$. Then the similarity $o(x_i)$ is concatenated with $f(x_i)$ and fed into the next layer of the discriminator. Minibatch discrimination penalizes the generator if mode is collapsing and encourages the model to generate diverse images. With minibatch discrimination, our cGAN models can generate more diverse images which can be effectively added to the training data for augmentation.

We then introduce the feature based filtering method after generating synthetic images from our trained cGAN model. One of the advantages of GANs is that the input is drawn from a distribution and one can generate an infinite number of images with a trained GAN. Thus, any filtering methods will not affect the number of images available to be added to the training data. We first pre-train a feature extractor using the baseline classifier on the original training set to extract visual features of the input images. The features consist of activations produced by different layers in the feature extractor.
The feature distance between image $x$ and centroid $c$ is then defined as 

\begin{small}
\begin{equation}
% \begin{split}
D_{f}(x,c) = \sum_{l} \frac{1}{H_lW_l} ||\hat{\phi_l}(x) - \hat{\phi_l}(c)||_{2}^{2} \enspace ,\label{Eq:feature}
% \end{split}
\end{equation}
\end{small}

\noindent where $\hat{\phi_l}$ is the unit-normalized activation in
the channel dimension $A_l$ of the $l$th layer of a feature extraction network with shape $H_l \times W_l$. This $\ell_2$ distance between unit-normalized activation can be regarded as a cosine distance in the feature space.

The centroid $c$ is calculated as the average feature of all training images in the same class. For class $i$, its centroid $c_i$ is represented by
\begin{small}
\begin{equation}
% \begin{split}
c_i = [\frac{1}{N_i}\sum_{j=1}^{N_i}\phi_1(x_j),\frac{1}{N_i}\sum_{j=1}^{N_i}\phi_2(x_j),...,\frac{1}{N_i}\sum_{j=1}^{N_i}\phi_L(x_j)] \enspace ,\label{Eq:centroid}
% \end{split}
\end{equation}
\end{small}
\noindent where $N_i$ denotes the number of training samples in $i$th class and $x_j$ is the $j$th training sample. Similar to Eq.~\ref{Eq:feature}, $\phi_l$ is the activation extracted from the $l$th layer of the feature extraction network. $L$ is the total number of layers utilized in the feature based filtering.
% \vspace{-5pt}
\subsection{Implementation Details}
% \vspace{-5pt}
Our GAN model is built upon the DCGAN~\cite{radford2015unsupervised} and WGAN-GP~\cite{gulrajani2017improved} with several alterations. The conditional generator consists of $8$ convolutional blocks, where the first block consists of two separate transpose convolutional blocks for conditional labels $c$ and random vector $z$. After the first block, the activations of $c$ and $z$ are concatenated along the channel dimension and fed into the next layer of the
generator. Different from DCGAN which uses transpose convolution for upsampling, our convolutional block consists of a bilinear upsampling with factor $2$ except for the first shared block which has factor $2\times1$, a $3\times3$ convolution with stride $1$, a batch normalization layer and a ReLU activation. The final block is a $7\times7$ convolution followed by a tanh activation.

The conditional discriminator consists of $7$ convolutional blocks, where the first block includes two separate convolutional blocks for conditional labels $c$ and random vector $z$ with no batch normalization as in DCGAN. The activations of $c$ and $z$ are concatenated and fed into the rest of convolutional blocks.
Each of the rest of the convolutional blocks contains a $4\times4$ convolutional layer, a batch normalization layer and a LeakyRelu activation. Each convolution has stride $2$ except for the $6$th and $7$th block which has stride $2\times1$ and $1$, respectively. Activations of the last convolutional block are then fed into the minibatch discrimination layer as described in Eq.~\ref{Eq:minibatch}. After the minibatch discrimination, a fully connected layer outputs the final logits of discriminator.

Our cGAN model is trained with WGAN-GP loss with batchsize $100$, fixed learning rate $2e-4$ and $700$ training epochs. The baseline classifier is the widely used ResNet18~\cite{he2016deep} which has shown promising performance in various vision tasks. All classification models are trained using the same baseline classification model with batchsize $64$ by Adam optimizer~\cite{kingma2014adam} with weight decay $2e-5$ and the cross-entropy loss. The initial learning rate of all classification models is $1e-3$ and is reduced by factor $0.2$ when the validation accuracy has stopped being improved for $5$ epochs. 
% TODO: add some details for baseline models.

\vspace{-10pt}
\section{Experiments}
\begin{figure}[t]
\begin{center}
  \includegraphics[width=0.99\linewidth]{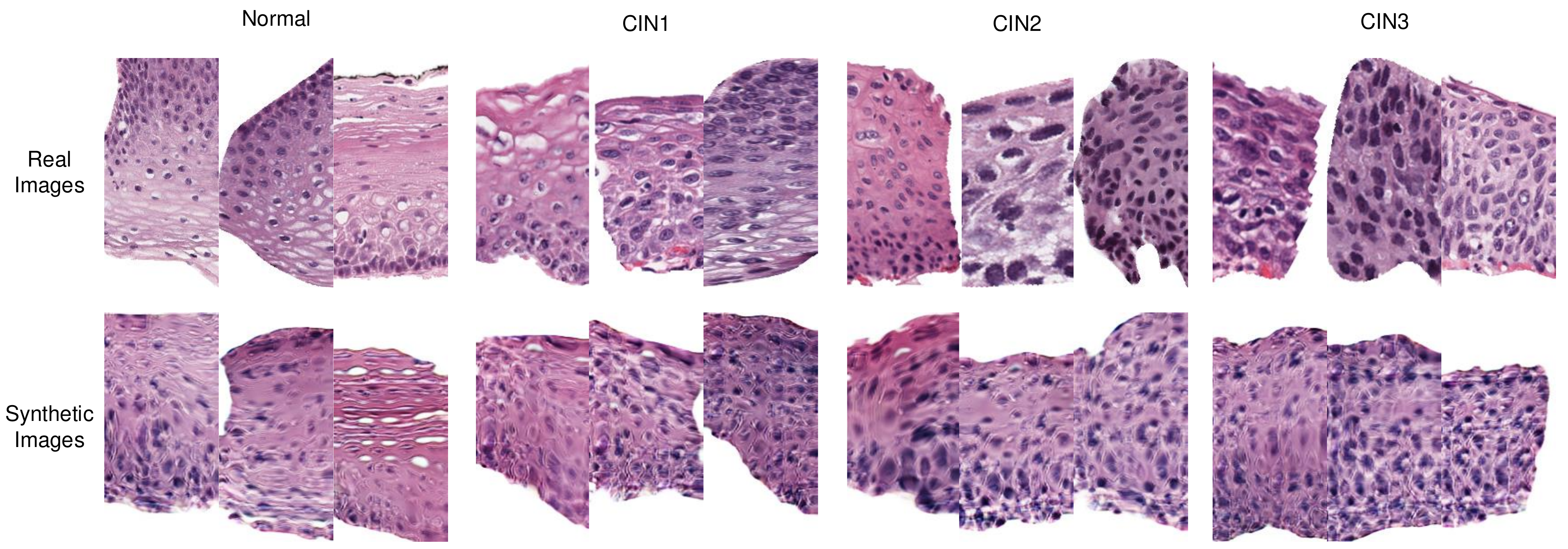}
\end{center}
  \caption{Examples of real and synthetic images for all CIN grades. }
  %The first row is real images in the original training set, the second row is cGAN generated images. From these examples, one can see that our models are capable of generating realistic cervical epithelium image patches.}
\label{fig:result}
% \vspace{-15pt}
\end{figure}

% \vspace{-5pt}
The experimental dataset is a cervical histopathology image dataset collected from a collaborating health sciences center. It contains multiple data sources, all of which are annotated by the same pathologist. The data processing follows~\cite{guo2016nuclei} by dividing an annotated image into patches according to the medial axis of the epithelium tissue in that image. We first divide the medial axis into several straight segments, and crop the patches with each containing one straight segment. The cropped patches are then rotated such that all their medial axes align horizontally after rotation. The cropped patches are resized to a unified size of $256 \times 128$. In total, there are $1,112$ normal, $181$ CIN1, $463$ CIN2, $454$ CIN3 patches. Examples of the images can be found in the first row of Fig.~\ref{fig:result}. All evaluations are done based on the patch-level ground truth annotations. We randomly split the dataset, by patients, into training, validation, and test sets, with ratio $7:1:2$ and keep the ratio of image classes almost the same among different sets. All evaluations and comparisons reported in this section are done on the test set.

We report quantitative evaluation scores between all baseline models and our models including the accuracy, area under the ROC curve (AUC), sensitivity and specificity to provide a comprehensive comparison. All models are run for $5$ rounds with random initialization for fair comparison. The mean and standard deviation results of $5$ runs are illustrated in Table~\ref{Tb:results}. We use the same baseline classifier with same hyperparameters setting in all experiments to ensure differences only come from the augmentation mechanisms.

We first conduct experiments with the baseline ResNet18~\cite{he2016deep} classifier trained on the original training data and the data with traditional data augmentation. The traditional data augmentation used in our experiments includes random horizontal and vertical flipping, random adjustment of brightness, contrast and saturation to serve as a color based augmentation. From the first two rows of Table~\ref{Tb:results}, one can see that there are no obvious improvements from leveraging traditional data augmentation, which further indicates that traditional augmentation methods do not perform well on such classification problem. Since we cannot ensure the feature quality of the traditional augmentation, we only train all our GAN models using the original training data.

Examples of the synthetic images generated by our cGAN models compared to the real images in the training set are shown in Fig.~\ref{fig:result}. As one can observe from Fig.~\ref{fig:result}, our generated images are realistic and keep important features such as color, shape and location of nuclei, and texture information so that the generated images can be used to extend the original training set. 

\begin{figure}[t]%
    \centering
    \subfloat[Expanded training data before filtering]{{\includegraphics[width=0.499\textwidth]{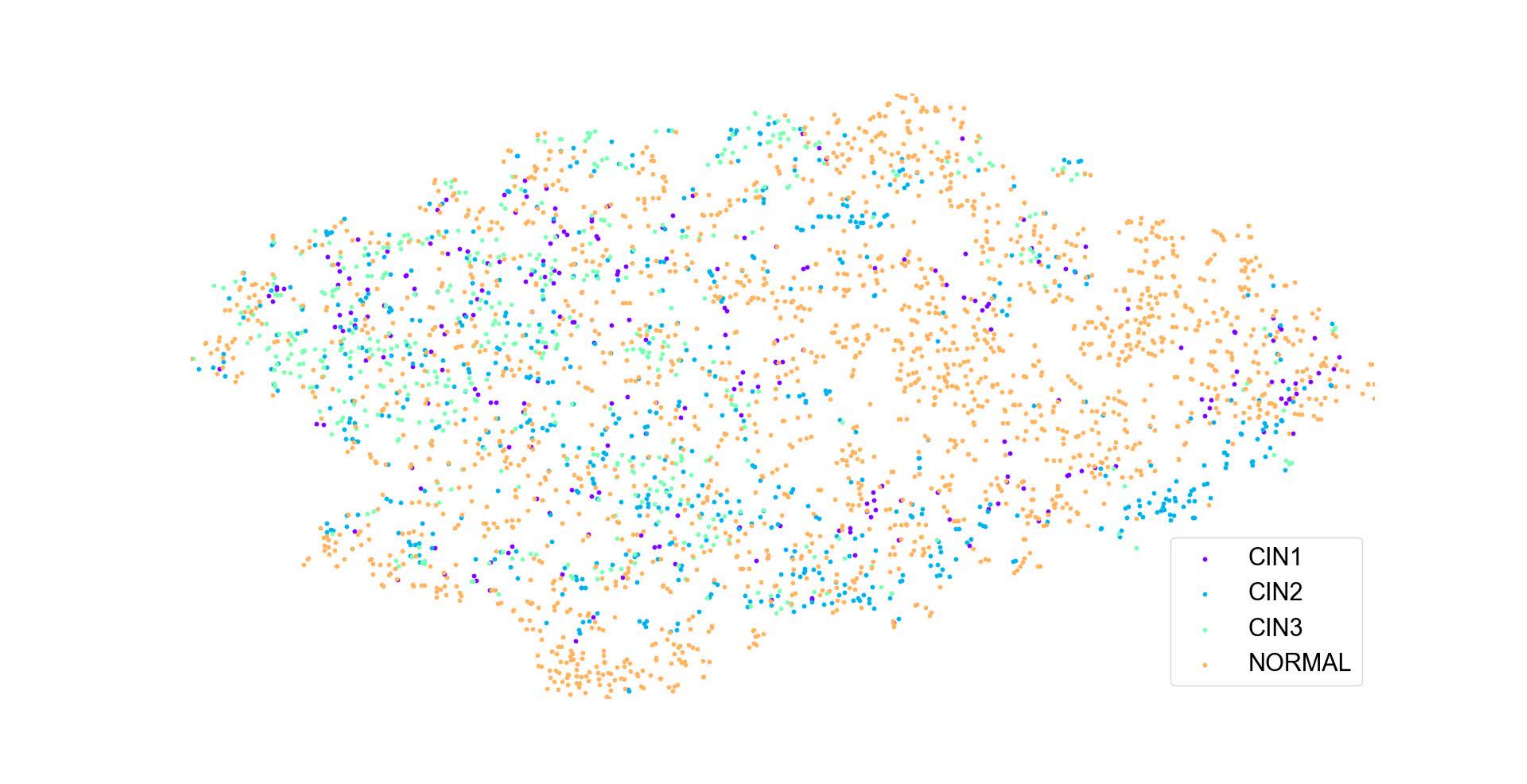} }}
    % \qquad
    \subfloat[Expanded training data after filtering]{{\includegraphics[width=0.499\textwidth]{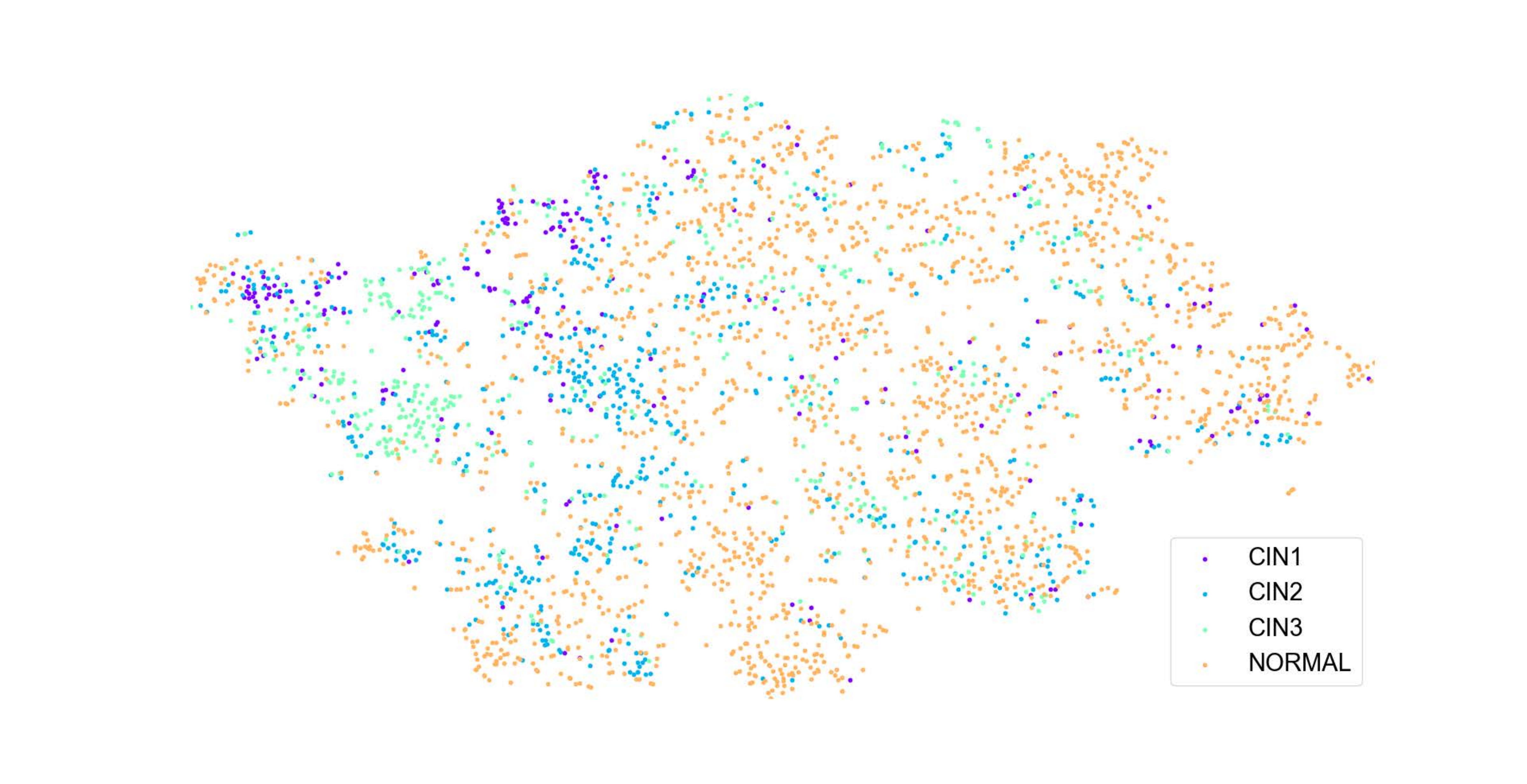} }}
    \caption{t-SNE visualization of extracted image features of expanded training data. The ratio of synthetic images to the original training images is set to 2 for both before and after filtering (R=2). }%
    \label{fig:features}%
% \vspace{-15pt}
\end{figure}

While the synthetic images are visually appealing, high realism is not equivalent to meaningful features for improving classification results. We further explore the distribution of generated images in the feature space in Fig.~\ref{fig:features} to ensure they are separable. We use the t-SNE~\cite{maaten2008visualizing} dimension reduction algorithm which can convert the embedding of high-dimensional data into a two-dimensional space for better visualization. After training a baseline ResNet18 classifier with the original training data, we use the pre-trained ResNet18 model as the feature extractor to extract features from the last convolutional layer in the ResNet18 model. We use the same feature extractor for both the expanded training data without and with feature based filtering in Fig.~\ref{fig:features} (a) and (b), respectively. Although overall there are overlapping between classes in the feature space which indicates the difficulty of such classification problem, the new training data with feature based filtering clearly have more distinguishable features than the ones without the feature filtering, which supports our claim and is in accord with the improvements in classification performance.

In all GAN-based augmentation models, the generated synthetic images are added into the original training set in equal proportion and we use ratio $R$ to represent the number of synthetic images to the number of real images. We keep the same proportion between classes as in the original set to make sure synthetic images generated from minority classes are meaningful for augmentation. We start with a cGAN model with an auxiliary classifier (AC)~\cite{odena2017conditional} which is similar to~\cite{frid2018gan}. The auxiliary classifier is added to the last layer of the discriminator network to output the class labels for both real and synthetic images. While cGAN with AC works fine when the ratio is $0.5$, the same model with ratio being $2$ significantly degrades the classification performance, resulting in even worse scores than the baseline model with no augmentation. One possible reason is that the auxiliary classifier is trained on both real and synthetic images (mainly trained on synthetic images when $R>1$), the features learned in the training process may not represent the original real data very well. Rather, if a feature extractor is trained only on the real data, the learned features could be more meaningful to the original data.

Following this idea, we provide an ablation study of our cGAN models with and without the feature filtering. For feature based filtering, we first generate $5,000$ images for each class, then the images with lowest distance to the corresponding centroids will be kept. Retained quantity is calculated based on the ratio $R$. As shown in Table~\ref{Tb:results}, our no filtering cGAN model with different ratios all show improvements in classification performance. Meanwhile, the feature filtering brings obvious benefits to all evaluation metrics, and our full models with different number of synthetic images added to the training data achieved best performance in all metrics. More importantly, the quality levels of generated images from our full models are very stable. Classification performances are similar among different ratios and during different runs, with consistently high mean and low std, which demonstrates that the feature qualities of generated images after filtering are superior to images generated by other models or without filtering. 

\begin{table}
% \vspace{-10pt}
\begin{center}
\begin{tabular}{l|c|c|c|c}
\hline
Method & Accuracy & AUC & Sensitivity & Specificity \\
\hline
ResNet18~\cite{he2016deep} & $0.663\pm0.032$ & $0.775\pm0.021$ & $0.553\pm0.019$ & $0.866\pm0.010$\\
\hline
Traditional Augmentation & $0.670\pm0.015$ & $0.780\pm0.010$ & $0.587\pm0.016$ & $0.868\pm0.006$ \\
\hline
cGAN + AC R=$0.5$ & ${0.687}\pm0.009$ & $0.792\pm0.006$ & $0.596\pm0.009$ & $0.874\pm0.003$ \\
\hline
cGAN + AC R=$2$ & $0.660\pm0.020$ & ${0.773}\pm0.014$ & ${0.551}\pm0.032$ & ${0.862}\pm0.011$ \\
\hline
% cGAN + AC w/ Filtering R=$0.5$ & $0.704\pm0.012$ & ${0.803}\pm0.008$ & ${0.615\pm0.019}$ & ${0.878}\pm0.004$ \\
% \hline
% cGAN + AC w/ Filtering R=$2$ & $0.649\pm0.021$ & ${0.766}\pm0.014$ & ${0.518}\pm0.033$ & ${0.855}\pm0.008$ \\
% \hline
Ours w/o Filtering R=$0.5$ & ${0.695}\pm0.009$ & $0.796\pm0.006$ & $0.559\pm0.032$ & $0.874\pm0.006$ \\
\hline
Ours w/o Filtering R=$2$ & $0.705\pm0.010$ & ${0.804}\pm0.007$ & ${0.555\pm0.019}$ & ${0.872\pm0.005}$ \\
\hline
Ours w/ Filtering R=$0.5$ & ${0.716}\pm0.009$ & ${0.810}\pm0.005$ & $\bm{0.611}\pm0.012$ & $\bm{0.886}\pm0.006$ \\
\hline
Ours w/ Filtering R=$2$ & $\bm{0.717}\pm0.008$ & $\bm{0.811}\pm0.006$ & ${0.608}\pm0.009$ & ${0.882}\pm0.003$ \\
\hline
\end{tabular}
\end{center}
\caption{Quantitative comparisons of baseline classification models and different augmentation models. All evaluation metrics are averaged over four classes. AC stands for the auxiliary classifier in the discriminator and R represents the data augmentation ratio.
%of how many number of new examples are added to the training data.   
}
\label{Tb:results}
% \vspace{-15pt}
\end{table}

% \vspace{-20pt}
\section{Conclusions}
% \vspace{-5pt}
In this paper, we investigate a novel GAN based aumentation pipeline for cervical histopathology image classification problem. We mainly focus on one of the major limitations of using GAN for augmentation: one cannot measure and control the quality of the synthetic images. While traditional GANs try to improve the fidelity of synthetic images, an augmentation system should try to generate images which have better feature quality rather than visual realism. By introducing a feature based filtering mechanism, our model boosts the performance of baseline classifier significantly on a challenging cervical histopathology dataset. As an attempt to make better use of GAN based augmentation models in medical imaging, our proposed pipeline has great potentials in other medical imaging applications where the number of labeled data is limited.

% \vspace{-10pt}
\section{Acknowledgements}
% \vspace{-5pt}
This research was supported in part by the Intramural Research Program of the National Institutes of Health (NIH), National Library of Medicine (NLM), and Lister Hill National Center for Biomedical Communications (LHNCBC), under Contract \# \\
HHSN276201800170P. We gratefully acknowledge the invaluable medical assistance from Dr. Rosemary Zuna, M.D. ,of the University of Oklahoma Health Sciences Center, and the work of Dr. Joe Stanley of Missouri University of Science and Technology which made the data collection possible.

% \vspace{-10pt}

% ---- Bibliography ----
%
% BibTeX users should specify bibliography style 'splncs04'.
% References will then be sorted and formatted in the correct style.
%

% \bibliographystyle{splncs04}
% {\small
% \bibliography{reference}
% }

\end{document}